\documentclass{article}
\usepackage{amsfonts,amssymb, color}
\usepackage{amsmath}
\usepackage{hyperref}
\usepackage{cleveref}
\usepackage{refcheck}
\usepackage{tikz}
\usepackage{tikz-cd}
\usepackage{caption}

\newtheorem{theorem}{Theorem}

\newtheorem{definition}[theorem]{Definition}
\newtheorem{example}[theorem]{Example}

\newtheorem{lemma}[theorem]{Lemma}

\newtheorem{proposition}[theorem]{Proposition}

\newtheorem{remark}[theorem]{Remark}

\newcommand{\ess}{\mathrm{ess}}
\newcommand{\esssup}{\ess\!\sup}

\newcommand{\di}{\diamond}

\newcommand{\C}{\mathcal{C}}

\newcommand{\E}{\mathbb{E}}
\newcommand{\F}{\mathcal{F}}
\newcommand{\G}{\mathbb{G}}
\newcommand{\Gcal}{\mathcal{G}}
\newcommand{\X}{\mathcal{X}}

\newcommand{\p}{\mathbb{P}}
\newcommand{\Q}{\mathbb{Q}}
\newcommand{\R}{\mathbb{R}}
\newcommand{\trho}{\tilde{\rho}}

\DeclareMathOperator{\dom}{dom}
\DeclareMathOperator{\epi}{epi}
\DeclareMathOperator{\gconv}{\otimes}
\DeclareMathOperator{\im}{Im}

\newcommand{\of}[1]{\ensuremath{\left( #1 \right)}}

\begin{document}

\title{\vspace{-2.2cm} On geometrically convex risk measures\thanks{This research was funded in part by the Netherlands Organization for Scientific Research under grant NWO Vici 2020--2027 (Ayg\"un, Laeven).}}
\author{M\"ucahit Ayg\"un 
\\
{\footnotesize Dept. of Quantitative Economics}\\
{\footnotesize University of Amsterdam}\\
{\footnotesize and Tinbergen Institute}\\
{\footnotesize \texttt{M.Aygun@uva.nl}}
\\
\and Fabio Bellini 
\\
{\footnotesize Dept. of Statistics and Quantitative Methods}\\
{\footnotesize University of Milano Bicocca}\\
{\footnotesize \texttt{Fabio.Bellini@unimib.it}}\\
\and Roger J.~A.~Laeven
\\
{\footnotesize Dept. of Quantitative Economics}\\
{\footnotesize University of Amsterdam, CentER}\\
{\footnotesize and EURANDOM}\\
{\footnotesize \texttt{R.J.A.Laeven@uva.nl}}
}

\date{\today}

\maketitle

\begin{abstract}
Geometrically convex functions
constitute an interesting class of functions obtained by replacing the arithmetic mean with the geometric mean in the definition of convexity. 
As recently suggested, geometric convexity may be a sensible property for financial risk measures (\cite{BLR18, LRG22, ABL23}). 

We introduce a notion of GG-convex conjugate, parallel to the classical notion of convex conjugate introduced by Fenchel, and we discuss its properties. 
We show how GG-convex conjugation can be axiomatized in the spirit of the notion of general duality transforms introduced in \cite{AM07, AM09}. 

We then move to the study of GG-convex risk measures, which are defined as GG-convex functionals defined on suitable spaces of random variables.\ 
We derive a general dual representation that extends analogous expressions presented in \cite{ABL23} under the additional assumptions of monotonicity and positive homogeneity.\ 
As a prominent example, we study the family of Orlicz risk measures.\
Finally, we introduce multiplicative versions of the convex and of the increasing convex order 
and discuss related consistency properties of law-invariant GG-convex risk measures. 
\end{abstract}


\section{Introduction}
While alternative notions of convexity based on geometric means have been discussed and analyzed in classical literature (\cite{M28,HLP52,RV73}), modern work almost exclusively restricts attention to the usual (arithmetic) convexity.
This applies not only to analysis, in particular to the whole body of convex analysis, but also to the wide variety of applications in which convexity plays a pivotal role.

Our aim in this note is to develop a general duality theory for GG-convex functionals, and discuss its application to the theory of financial risk measures. 
Geometric convexity is a variant of the usual definition of convexity, in which the arithmetic mean is replaced by the geometric mean as follows: a function $f \colon (0,\infty) \to (0,\infty$) is said to be geometrically convex (GG-convex, for short) if for each $x,y>0$ and $\lambda \in (0,1)$ it holds that
\[
f(x^\lambda \cdot y^{1-\lambda})\leq f^\lambda(x) \cdot f^{1-\lambda}(y).
\]
\noindent
From a purely graphical point of view, GG-convex functions are positive functions of a positive variable that are convex when plotted on a log-log graph.\ 
Equivalently, GG-convex functions are log-convex functions of $\log x$; in the literature GG-convex functions have also been called multiplicatively convex or log-log convex.\ 
We refer to \cite{N00, NP04} and the references therein for general background on GG-convex functions and to \cite{ADB19} for novel recent applications in optimization.\ 
As remarked in \cite{N00}, ``nowadays, the subject of multiplicative convexity seems to be even forgotten, which is a pity because of its richness''. 

We begin by reviewing the relevant properties of GG-convex functions and their relations with other notions of convexity. 
As in the case of convex functions, it is useful to consider functions admitting $0$ and $\infty$ as possible values. 

Our first novel contribution is the introduction of a notion of GG-convex conjugate for real functions of a real variable defined by the following formula:
\[
f^\di (y) :=\sup_{x > 0} \left \{ \frac{\exp ( \log x \log y)}{f(x)} \right \}. 
\]
\noindent
GG-convex conjugation is a multiplicative version of the Fenchel transform and has similar properties, such as being \emph{order-reversing} in the sense that 
$
f \leq g \Rightarrow f^\di \geq g^\di, 
$
\emph{involutive} in the sense that under suitable assumptions
$
f^{\di \di} = f,
$
and \emph{multiplicative with respect to multiplicative inf-convolution} in the sense that
$
(f \gconv g)^\di = f^\di \cdot g^\di,
$
where
$
(f \gconv g)(z) := \inf_{x,y>0,\, xy=z} \left \{f(x) \cdot g(y) \right \}. 
$

Remarkably, these properties are also sufficient to characterize the general structure of GG-convex transforms up to a few additional parameters, and they indeed form the basis of two possible axiomatizations of GG-convex transform presented in Theorems~\ref{th:GG-axiom} and \ref{th:GG-axiom-2}, following the spirit of \cite{AM07, AM09}, showing that GG-convex conjugation is a general duality transform in the sense of \cite {AM09}. 

We then move to the study of GG-convex functionals on suitable function spaces, interpreted as financial risk measures on general domains as in \cite{BKMMS21}. 
The financial relevance of the notion of GG-convexity for the study of return risk measures has been discussed in several recent papers (see e.g., \cite{ABL23, BLR18, LRG22}). 

Our main result here is the following dual representation of GG-convex risk measures satisfying a version of the Fatou property, given in Proposition~\ref{prop:GG-duality-2}:
\begin{equation*}
	\rho (X)=\sup_{Y \in \X^\ast_{\rm{log}} } \left \{ \frac{\exp ( \E [\log Y \log X])}{\rho^\di (Y)} \right \}. 
\end{equation*}
\noindent
We then discuss how similar dual representations given in \cite{ABL23} for return risk measures, i.e., under the additional assumptions of positive homogeneity, monotonicity and normalization can be derived as special cases of Proposition~\ref{prop:GG-duality-2}. 

Finally, we focus on the law-invariant case and we show that law-invariant GG-convex risk measures are consistent with respect to a logarithmic version of the convex order, in the sense that if $\rho$ is GG-convex and law-invariant, then
\[
\log X \leq_{\rm{cx}} \log Y \Rightarrow \rho(X) \leq \rho(Y). 
\] 

The remainder of this note is structured as follows. 
Section~\ref{sec:GG} briefly reviews relevant results on geometric convexity.\ 
Section~\ref{sec:GG-dual-1} introduces duality theory for GG-convex real functions and establishes their axiomatization.\ 
Section~\ref{sec:GG-RM} presents duality theory for GG-convex risk measures.\ 
Section~\ref{sec:GG-Order} introduces the relevant stochastic orders and discusses the corresponding consistency properties.\ 
Section~\ref{sec:con} provides a short conclusion.\
All proofs are moved to the Appendix. 

\section{Geometric convexity}\label{sec:GG}
We start by recalling the notion of geometric convexity (also called log-log convexity in the literature, and referred to as GG-convexity throughout the paper) for real functions of a single variable, possibly taking the values $0$ and $\infty$.  
\begin{definition}
\label{def:GG-conv}
A function $f \colon (0,\infty) \to [0, \infty]$ is called GG-convex if for each $x,y \in (0,\infty)$ and $\lambda \in (0,1)$ it holds that
\begin{equation}\label{eq:GG-conv}
f(x^\lambda \cdot y ^{1-\lambda}) \leq  f^\lambda(x) \cdot f^{1-\lambda} (y), 
\end{equation}
where we set by definition $0 \cdot \infty = \infty$. The \emph{effective domain} of $f$ is 
\[
\dom f = \{x \in (0,\infty) \mid f(x) < \infty \},
\]
and $f$ is said to be \emph{proper} if $f(x)>0$ for each $x \in (0,\infty)$ and $\dom f \neq \emptyset$. 
\end{definition}
\noindent
We refer to \cite{N00, NP04} for the most important properties of GG-convex functions, although only the finite case is considered in these references.\ 
Notice that $\dom f$ is convex and that $f(x)=0,\, f(y)<\infty \Rightarrow f(z)=0$ for each $z \in [x,y)$. 
We give a few examples that will play an important role in the following. 

\begin{example}
Let $C \subseteq (0,\infty)$ and let $\delta_C$ be the indicator function of $C$, i.e.,
\[
\delta_C :=
\begin{cases}
0 &\text {if } x \in C \\
\infty &\text {if } x \not \in C. 
\end{cases}
\]
Then $\delta_C$ is GG-convex if and only if $C$ is convex. 
\end{example}

\begin{example}
Let $f(x)=Ax^B$, with $A>0$ and $B \in \R$. 
Then $f$ is GG-convex, and as we will see in Definition~\ref{def:GG-conc} below $f$ is also GG-concave, so functions of this kind will be referred to as \emph{GG-affine}.
\end{example}

\begin{example}
Let $f$ be a polynomial with nonnegative coefficients or more generally a real analytic function whose series expansion has nonnegative coefficients. 
Then $f$ is GG-convex, taking the value $\infty$ outside its disk of convergence (see \cite{N00}, Proposition~2.4). 
\end{example}

\noindent
For the sake of completeness, we recall a few basic properties of GG-convex functions in Proposition~\ref{prop:GG-prop} below.\ 
Letting $X \colon (\Omega, \F, \p) \to (0,\infty)$ be a positive random variable such that $\log X \in L^1(\Omega, \F, \p)$, we denote with $\G[X]$ the geometric mean of $X$, which is defined by $\G[X]=\exp \left ( \E[ \log X] \right)$.

\begin{proposition}
\label{prop:GG-prop}
\begin{itemize}
\item []
\item [a)] $f,g$ are GG-convex $\Rightarrow$ $f+g$ and $f\cdot g$ are GG-convex.
\item [b)] $f_\alpha$ is GG-convex for each $\alpha \in I$ $\Rightarrow$ $\sup_{\alpha \in I} f_\alpha$ is GG-convex.
\item [c)] $f$ is GG-convex if and only if there exists $g \colon (-\infty, \infty) \to [-\infty, \infty]$ convex such that 
$
f=\exp \circ \,  g \circ \log ,
$
where we set by definition $\exp(-\infty)=0$, $\exp(\infty)=\infty$.\ 
Further, $f$ is nondecreasing if and only if $g$ is nondecreasing. 
\item [d)] if $f$ is GG-convex then for each positive random variable $X$ 
such that $\log X \in L^1(\Omega, \F, \p)$ it holds that
$
f(\G[X]) \leq \G[f(X)]
$.
\item [e)] If $f$ is finite and $f \in C^2(0,\infty)$ then $f$ is GG-convex if and only if
\[
x \{ f''(x)f(x)-[f'(x)]^2\}  + f(x)f'(x) \geq  0, \, \forall x \in (0,\infty). 
\]
\end{itemize}
\end{proposition} 
Since constant functions are trivially GG-convex, item a) implies that GG-convex functions form a convex cone.\ 
Item c) gives a correspondence between GG-convex and convex functions that will be used many times in the paper.\ 
Item d) is a variant of Jensen's inequality for GG-convex functions. 
It will be useful to introduce also the following two closely related notions of convexity.
\begin{definition}
\label{def:AG-GA-conv}
A function $f \colon (-\infty, \infty) \to [0,\infty] $ is called AG-convex if for each $x,y \in \R $ and $\lambda \in (0,1)$ it holds that
\[
f(\lambda x + (1-\lambda)y) \leq f^\lambda(x) \cdot f^{1-\lambda}(y),
\]
where we set by definition $0 \cdot \infty = \infty$.\ AG-convex functions are often also called log-convex, since by definition $f$ is AG-convex if and only if $\log f$ is convex. \smallskip \\
\noindent
A function $f \colon (0,\infty) \to [-\infty, \infty] $ is called GA-convex if for each $x,y \in (0,\infty) $ and $\lambda \in (0,1)$ it holds that
\[
f(x^\lambda \cdot y^{1-\lambda}) \leq \lambda f(x) + (1-\lambda) f(y),
\]
where we set by definition $- \infty + \infty = \infty$.\ 
A function $f$ is GA-convex if and only if $f(e^x)$ is convex. 
\end{definition}

The definition of AG-convexity requires a comparison between an Arithmetic and a Geometric mean, while GA-convexity involves a Geometric and an Arithmetic mean.\ 
We will refer to the usual convexity as AA-convexity and collectively to these notions as ``algebraic convexities'', following e.g.,\ \cite{NP04}. 

We report in the following lemma a few implications immediately following from the AM-GM inequality.
The proof is elementary and thus omitted. 
\begin{lemma}
\label{lem:AM-GM}
\begin{itemize}
\item[]
\item [a)] $f$ AG-convex $\Rightarrow f$ AA-convex
\item [b)] $f$ GG-convex $\Rightarrow f$ GA-convex
\item [c)] $f$ nondecreasing and AA-convex $\Rightarrow f$ GA-convex
\item [d)] $f$ nondecreasing and AG-convex $\Rightarrow f$ GG-convex
\item [e)] $f$ nondecreasing and AG-convex, $g$ GA-convex $\Rightarrow f \circ g$ GG-convex
\item [f)] $f$ nondecreasing and GG-convex, $g$ GG-convex $\Rightarrow f \circ g$ GG-convex.
\end{itemize}
\end{lemma}

\noindent The relationships holding among the various classes of algebraically convex functions in the nondecreasing case are represented in the diagram in Fig.\ 1.

\begin{equation*}
\begin{array}{c@{}c@{}c}
AA & \implies & GA \\
\big \Uparrow & & \big \Uparrow  \\
AG & \implies & GG
\end{array}
\end{equation*}
\captionof{figure}{Implications among nondecreasing algebraically convex functions.}
\smallskip

\noindent All the arrows are one-sided, as the following elementary examples show.
\begin{example}
\begin{itemize}
\item []
\item [i)] Let $f(x)=e^x$.\ Then $f \in AG, AA, GG, GA$.
\item [ii)]  Let $f(x)=x^2$.\ Then $f \in AA, GG, GA$ but $f \not \in AG$.
\item [iii)] Let $f(x)= \log x$.\ Then $f \in GA$, but $f \not \in AG,AA,GG$.
\end{itemize}
\end{example}

\noindent GG-concave functions are defined just by reversing the inequality in \eqref{eq:GG-conv}. 
\begin{definition}
\label{def:GG-conc}
A function $f \colon (0,\infty) \to [0, \infty]$ is called GG-concave if for each $x,y \in (0,\infty)$ and $\lambda \in (0,1)$ it holds that
\[
f(x^\lambda \cdot y ^{1-\lambda}) \geq  f^\lambda(x) \cdot f^{1-\lambda} (y), 
\]
where we set by definition $0 \cdot \infty = 0$. 
The \emph{effective domain} of $f$ is 
\[
\dom f = \{x \in (0,\infty) \mid f(x) > 0 \},
\]
and $f$ is \emph{proper} if $f(x) < \infty$ for each $x \in (0,\infty)$ and $\dom f \neq \emptyset$. 
\end{definition}
The properties of GG-concave functions can be derived from those of GG-convex functions by noticing that 
$f$ is GG-concave if and only if $1/f$ is GG-convex. 

%

\section{GG-convex duality}\label{sec:GG-dual-1}
The first main contribution of the paper is the introduction of a notion of conjugation for GG-convex functions, similar to the well-known Fenchel conjugate. 
 
\begin{definition}\label{def:GG-conjugate}
Let $f \colon (0,\infty) \to [0,\infty]$.\ The \emph{GG-convex conjugate} of $f$ is
\[
f^\di (y) :=\sup_{x > 0} \left \{ \frac{\exp ( \log x \log y)}{f(x)} \right \}, 
\]
where we set by convention $1/0=\infty$ and $1/\infty=0$. 
\end{definition}
\noindent Notice that equivalently
$$
f^\di (y) = \sup_{x > 0} \left \{ \frac{x^{\log y}}{f(x)} \right \}= \sup_{x > 0} \left \{ \frac{y^{\log x}}{f(x)} \right \}.
$$
Further, if $f(\bar{x})=0$ for some $\bar{x} \in (0,\infty)$, then $f^\di (y)= \infty$ for each $y \in (0,+\infty)$, while if $f(x)=\infty$ for each $x \in (0,\infty)$, then $f^\di=0$.\ 
It is possible to give a geometric interpretation to the GG-convex conjugate that closely resembles the one of the Fenchel convex conjugate. Recall the definition of epigraph: 
\[
\epi f =\{(x,y) \mid x \in \dom f, \, y \geq f(x) \}. 
\]
For $\alpha >0$ it holds that
\[
f^\di(y) \leq \alpha \iff \frac{\exp ( \log x \log y)}{f(x)}\leq \alpha \iff f(x) \geq \frac{x^{\log y}}{\alpha},
\]
so points of $\epi f^\di$ parametrize GG-affine functions majorized by $f$, exactly as in the Fenchel case, where the points of $\epi f^\ast$ parametrize AA-affine functions majorized by $f$.\ 
Further, it holds that $f^\di(y) = \inf \{\alpha >0 \mid f(x) \geq  x^{\log y}/ \alpha \}$.

We illustrate GG-convex conjugation by means of a few elementary examples. 

\begin{example}
Let $f(x)=\delta_A + B$, with $A>0, B > 0$.\ 
Then 
$
f^\di(y) = y^{\log A}/B.
$
Conversely, if $f(x)=Ax^B$, with $A>0, B \in \R$ then 
$
f^\di(y) = 1/A + \delta_{\exp(B)}, 
$
so GG-affine functions and indicator functions are GG-conjugates of each other. 
\end{example}

\begin{example}
Let  $f(x)=e^x$.\ 
Then
\[
f^{\di}(y)=
\begin{cases}
1 &\text{ if } y \leq 1 \\
\frac{(\log y)^{\log y}}{y} &\text{ if } y > 1
\end{cases}
\]
since for $y>1$ the supremum is achieved in $x^\ast=\log y$. 
\end{example}

\begin{example} Let
$f=1+ \delta_{A}$,  $g=1+ \delta_{B}$, with $A,B >0$, $A \neq B$. 
Then
$0=\left [ \sup (f,g) \right]^\di < \inf (f^\di, g^\di) = \inf (y^{\log A}, y^{\log B})$,  
with a strict inequality in d). 
\end{example}

The main properties of the GG-convex conjugation are collected in the following proposition. 
Recall that a function $f$ is said to be lower semicontinuous if $\epi f$ is closed.\ 
We refer to e.g.,\ \cite{R70, RV73, Z02} for this notion and for background material on convex duality. 
\begin{proposition}\label{prop:GG-conjugate}
Let $f\colon (0,+\infty) \to (0,+\infty)$ and let $f^\di$ be as in Definition \ref{def:GG-conjugate}.\
\begin{itemize}
\item [a)] $f^\di$ is GG-convex and lower semicontinuous
\item [b)]  $f \leq g  \Rightarrow f^\di \geq g^\di$
\item [c)] $\left [\inf_{\alpha \in I} f_\alpha \right]^\di = \sup_{\alpha \in I} f_\alpha^\di$
\item [d)]  $\left[ \sup_{\alpha \in I} f_\alpha \right]^\di \leq \inf_{\alpha \in I} f_\alpha^\di$
\item [e)]  $[Af]^\di = f^\di/A$, for each $A>0$
\item [f)] $ [f(Ax)]^\di (y) = f^\di(y) \cdot y ^{-\log A}$, for each $A>0$
\item [g)] $[f(x^A)]^\di (y) = f^\di (y^{1/A})$,  for each $A \in \R$
\item [h)] $[f(x) \cdot x^A]^\di (y) = f^\di (y/e^A)$,  for each $A \in \R$.
\end{itemize}
\end{proposition}

A basic result in convex duality is the Fenchel-Moreau theorem, giving a sufficient condition for the equality of a function with its double conjugate.\ 
We present below a version of this result for GG-convex functions. 

\begin{proposition} \label{prop:GG-duality}
Let $f \colon (0,\infty) \to [0,\infty]$.\ 
Then $f^{\di \di} \leq f$.\ 
If $f$ is GG-convex, proper and lower semicontinuous (lsc) then $f^{\di \di}=f$,
that is,
\[
f(x)=\sup_{y > 0} \left \{ \frac{\exp ( \log x \log y)}{f^\di (y) }\right \}. 
\]
\end{proposition}
\noindent

Letting now
\[
\Gcal (0,\infty) := \left \{ f \colon (0,\infty) \to [0, \infty] \mid f \text{ is GG-convex, proper and lsc} \right \}, 
\]
Proposition~\ref{prop:GG-duality} shows that GG-convex conjugation, seen as a transform $^\di \colon \Gcal \to \Gcal$, is injective, surjective and involutive, in the sense that $f^{\di \di}=f$.\ 
Further, as we have seen in Proposition~\ref{prop:GG-conjugate}, it is also order-reversing, in the sense that $f \leq g \iff f^\di \geq g^\di$.\ 
In the same spirit of the general duality theory developed in \cite{AM07, AM09}, we show in the following theorem that these two properties are actually sufficient for axiomatizing the general structure of the $^\di$-transform.

\begin{theorem}\label{th:GG-axiom}
A one-to-one function $T \colon \Gcal(0,\infty) \to \Gcal(0,\infty)$ satisfies
\begin{itemize}
\item [a)] $T \left (T(f) \right)=f$ for each $f \in \Gcal(0,\infty)$ 
\item [b)] $f\geq g \Rightarrow T(f)\leq T(g)$ for each $f,g \in \Gcal(0,\infty)$ 
\end{itemize}
if and only if
\begin{equation}\label{Ti-ABC}
(T(f))(x) = Ax^{\log B} f^\di(B x ^C), 
\end{equation}
for some $A>0$, $B>0$, $C \in \R$. 
\end{theorem}

One of the most important properties of the usual convex conjugation is its additivity with respect to the operation of additive inf-convolution of functions, defined by $(f \oplus  g)(x)= \inf_{x_1 + x_2=x} \left \{f(x_1) + g(x_2) \right \}$.\ 
Similarly, we show in Proposition~\ref{prop:g-conv} below that the operation of GG-convex conjugation is multiplicative with respect to the following notion of multiplicative inf-convolution. 

\begin{definition}\label{def:g-conv}
Let $f, g \colon (0, +\infty) \to [0,+\infty]$.\ We define the multiplicative inf-convolution of $f$ and $g$ by
\[
(f \gconv g)(x) = \inf_{x_1 \cdot x_2=x} \left \{f(x_1) \cdot g(x_2) \right \}, 
\]
where we set by definition $0 \cdot \infty = \infty$. 
\end{definition}

To ease the notation and clarify the relation between additive and multiplicative inf-convolution, we set
\[
\C (\R) := \left \{ f \colon \R \to \R \mid f \text{ is proper, convex and lsc} \right \},
\]
and for each $f \in \Gcal (0,\infty)$ we define $C(f) \in \C(\R)$ by
$
C(f) := \log \circ \, f \circ \exp,
$
where it follows from Proposition~\ref{prop:GG-prop} that $C \colon \Gcal (0,\infty) \to \C(\R)$ is one-to-one. 

\begin{proposition}\label{prop:g-conv}
Let $f, g \colon (0,+\infty) \to [0,+\infty]$ and let $f \gconv g$ be as in Definition \ref{def:g-conv}.\ Then  
\[
(f \gconv g)^\di = f^\di \cdot g^\di. 
\]
Further, for each $f, g \in \Gcal (0,\infty)$, it holds that
\[
C (f \otimes g) =C(f) \oplus C(g),
\]
and, for each $f, g \in \C (\R)$, it holds that
\[
C^{-1} (f \oplus g) =C^{-1}(f) \otimes C^{-1}(g). 
\]
\end{proposition}

Another possible axiomatization of the $^\di$-transform among involutive transforms, based on its multiplicativity property with respect to multiplicative inf-convolutions, is given in the following theorem.  

\begin{theorem}\label{th:GG-axiom-2}
	A one-to-one transformation $T \colon \Gcal(0,\infty) \to \Gcal(0,\infty)$ satisfies
	\begin{itemize}
		\item [a)] $T(T(f))=f$ for each $f \in \Gcal(0,\infty)$ 
		\item [b)] $T(f \gconv g) = T(f) \cdot T(g)$
	\end{itemize}
	if and only if
	\begin{equation}\label{Ti-A}
		(T(f))(x) =  f^\di(x^A), 
	\end{equation}
	for some $A \in \R$. 
\end{theorem}

\section{GG-convex risk measures}\label{sec:GG-RM}

We now move to the study of GG-convex risk measures, defined as GG-convex functionals on spaces of random variables.\ We follow the setting of \cite{BKMMS21}.\ Let $(\Omega, \F, \p)$ be a nonatomic probability space.\ A linear subspace $\X \subseteq L^0(\Omega, \F, \p)$ is law-invariant if $X \in \X,\, Y \overset{d}{=} X \Rightarrow Y \in \X$, where $\overset{d}{=}$ denotes equality in distribution. We let $\X, \X^\ast$ be linear subspaces of $L^0(\Omega, \F, \p)$ such that:
\begin{itemize}
\item [i)] $\X$ and $\X^\ast$ are law-invariant
\item [ii)] $XY \in L^1(\Omega, \F, \p)$, for each $X \in \X$ and $Y \in \X^\ast$
\item [iii)] $L^\infty (\Omega, \F, \p) \subseteq \X \subseteq L^1 (\Omega, \F, \p)$ and $L^\infty (\Omega, \F, \p) \subseteq \X^\ast \subseteq L^1 (\Omega, \F, \p)$.
\end{itemize}
As discussed in \cite{BKMMS21}, this setting is general enough to encompass the most common rearrangement-invariant functions spaces, such as $L^p$ spaces and Orlicz spaces. 
Further, we set
\begin{align*}
\X_{\rm{log}} &= \{ X \in L^0(\Omega, \F, \p) \mid \log X \in \X \}, \\
\X^\ast_{\rm{log}} &= \{ X \in L^0(\Omega, \F, \p) \mid \log X \in \X^\ast \}.
\end{align*}
Similar to Definitions~\ref{def:GG-conv} and \ref{def:AG-GA-conv} we give the corresponding definitions of the various notions of convexity in the case of risk measures.
\begin{definition}
\begin{itemize}
\item[]
\item [a)] $\rho \colon \X_{\rm{log}} \to [0,\infty]$ is GG-convex if for each $X,Y \in \X_{\rm{log}}$ 
and $\lambda \in (0,1)$ 
\begin{equation*}
\rho (X^\lambda Y^{1-\lambda}) \leq  \rho^\lambda(X) \cdot \rho^{1-\lambda} (Y), 
\end{equation*}
where we set by definition $0 \cdot \infty = \infty$. 
\item [b)] $\rho \colon \X_{\rm{log}} \to [-\infty,\infty]$.\
is GA-convex if for each $X,Y \in \X_{\rm{log}}$ 
and $\lambda \in (0,1)$ 
\begin{equation*}
\rho (X^\lambda  Y^{1-\lambda}) \leq  \lambda \rho(X) + (1-\lambda) \rho(Y),
\end{equation*}
where we set by definition $\infty - \infty = \infty$.
\item [c)] $\rho \colon \X \to [0,\infty]$ is AG-convex if for each $X,Y \in \X$ 
and $\lambda \in (0,1)$ 
\begin{equation*}
\rho (\lambda X + (1-\lambda) Y) \leq  \rho^{\lambda}(X) \cdot \rho^{1-\lambda}(Y),
\end{equation*}
where we set by definition $0 \cdot \infty = \infty$. 
\end{itemize}
\noindent
\end{definition}
Two simple GG-convex functionals are given in the following example.
\begin{example}
Let $f \colon (0,+\infty) \to (0,+\infty)$ be GG-convex.
Then $\rho(X)=\E[f(X)]$ is GG-convex since
\[
\rho(X^\lambda Y^{1-\lambda}) = \E [f(X^\lambda Y^{1-\lambda})] \leq \E[[f(X)]^\lambda [f(Y)]^{1-\lambda}] \leq (\E[ f(X)])^\lambda \cdot (\E[ f(X)])^{(1-\lambda)}.
\]
Similarly, also $\rho(X)= \G[f(X)]$ is  GG-convex since
\[
\rho(X^\lambda Y^{1-\lambda}) = \G [f(X^\lambda Y^{1-\lambda})] \leq \G[[f(X)]^\lambda [f(Y)]^{1-\lambda}] = \rho^\lambda(X) \cdot \rho^{1-\lambda}(Y).
\]
\end{example}
\noindent
\smallskip
We further recall that a risk measure is said to be monotone if 
\[
X \leq Y \Rightarrow \rho(X) \leq \rho(Y),
\]
positively homogeneous if 
\[
\rho(\lambda X) = \lambda \rho(X) \text { for each } \lambda >0,
\]
and normalized if $\rho(1)=1$. 
Monotone, positively homogeneous and normalized risk measures have been called return risk measures in \cite{BLR18, LRG22, ABL23}.\ 
A risk measure is law-invariant if $X \overset{d}{=}Y \Rightarrow \rho(X)=\rho(Y)$.\ 
Similar to Lemma~\ref{lem:AM-GM}, we have the following implications among the various classes of algebraically convex risk measures.  

\begin{lemma}\label{lem:AM-GM-2}
\begin{itemize}
\item []
\item [a)] $\rho \text { AG-convex } \Rightarrow \rho \text { AA-convex}$
\item [b)] $\rho \text { GG-convex } \Rightarrow \rho \text { GA-convex}$
\item [c)] if $\rho$ is monotone, then $\rho \text { AA-convex } \Rightarrow \rho \text { GA-convex}$
\item [d)]  if $\rho$ is monotone, then $\rho \text { AG-convex } \Rightarrow \rho \text { GG-convex}$
\item [e)]  if $\rho$ is positively homogeneous, then $\rho \text { GA-convex } \Rightarrow \rho \text { GG-convex}$.
\end{itemize} 
\end{lemma}

The relationships holding among these classes in the case of return risk measures can be visualized by means of the following diagram. 
In comparison with Figure 1, we now have that GG- and GA-convexity are equivalent. 

\[
\begin{array}{c@{}c@{}c}
AA & \implies & GA \\
\big \Uparrow & & \big \Updownarrow  \\
AG & \implies & GG
\end{array}
\]
\captionof{figure}{Implications among classes of monotone, positively homogeneous and normalized algebraically convex risk measures.} 
\medskip


In the examples below, we provide a collection of GG-convex return risk measures. 
\begin{example}
\label{ex:Gmean}
Let $\rho(X)= \G[X]$. 
Then $\rho$ is a law-invariant GG-convex return risk measure that is not AA-convex. 
\end{example}

\begin{example}
\label{ex:pnorm}
Let $\rho(X)= \Vert X \Vert_p$, with $0<p<1$.\ 
Then $\rho$ is a law-invariant GG-convex return risk measure since from the H\"older inequality
\[
\rho(X^\lambda Y ^{1-\lambda})= \E \left [ X^{\lambda p} \cdot Y^{(1-\lambda)p} \right]^{1/p} \leq \E[X^p]^{\lambda p}  \cdot \E [Y^p]^{(1-\lambda) p}  = \rho(X)^{\lambda} \cdot \rho(Y)^{1-\lambda}.
\]
As in the previous example, $\rho$ is not AA-convex. 
\end{example}

\begin{example}
Let $\mathbf{P} = \{ \Q \text { prob.\ meas.\ on } (\Omega, \F) \mid \Q \ll \p \}$ and let $\mathcal{M} \subseteq \mathbf{P}$. For $\Q \in \mathbf{P}$, let $\G_\Q[X] := \exp \left (  \E_\Q[\log X] \right)$.\ Then 
\[
\rho(X)= \sup_{\Q \in \mathcal{M}} \G_\Q[X] 
\]
is a GG-convex return risk measure, that is law-invariant if $\mathcal{M}$ is law-invariant in the sense that
\[
\Q_1 \in \mathcal{M},\, \frac{d\Q_1}{d\p} \overset{d}{=} \frac{d\Q_2}{d\p} \Rightarrow \Q_2 \in \mathcal{M}. 
\]
\end{example} 

\begin{example}
In \cite{ABL23} the following class of risk measures has been considered. 
Let $\mathbf{P}$ be as in the previous example and let 
$\alpha \colon \mathbf{P} \to [0,1]$. Let
\[
\rho(X)= \sup_{\Q \in \mathbf{P}} \{ \alpha(\Q) \G_\Q[X] \}. 
\]
Then $\rho$ is a GG-convex return risk measure, which is law-invariant if $\alpha$ is law-invariant in the sense that 
\[
\frac{d\Q_1}{d\p} \overset{d}{=} \frac{d\Q_2}{d\p} \Rightarrow \alpha(\Q_1) = \alpha (\Q_2). 
\] 
\end{example}

\begin{example}
In \cite{BLR18} it was shown that in general if $\tilde{\rho}$ is a monetary convex risk measure, then
\[
\rho(X) = \exp \left ( \tilde{\rho} (\log X) \right)
\]
is a GG-convex return risk measure, which is law-invariant if $\tilde{\rho}$ is law-invariant. 
\end{example}


An interesting family of risk measures that illustrates well the various types of algebraic convexity are Orlicz risk measures, usually called Orlicz premia in the actuarial literature.\ 
We refer to \cite{HG82, BRG08, ABL23} and the references therein for general background on Orlicz premia. 
We recall their definition as given in \cite{ABL23}.

\begin{definition}
Let $\Phi \colon (0, \infty) \to \R$ satisfy:
\begin{itemize}
\item [a)] $\Phi(1)=1$, $\lim_{x \to \infty}\Phi(x)=\infty$
\item [b)] $\Phi$ is nondecreasing
\item [c)] $\Phi$ is left-continuous
\end{itemize}
For $X \in L^{\infty}_{+}$, the Orlicz premium is defined by
\[
H_\Phi(X) = \inf \{k > 0 \mid  \E \left [\Phi(X/k ) \right] \leq 1 \}.
\]
\end{definition}
Notice that this definition is more general than the classic one since we do not require that $\Phi(0)=0$ and that $\Phi$ is positive. 
As discussed in \cite{ABL23}, this allows the inclusion of interesting examples such as expectiles and the geometric mean itself. 
It is well known that Orlicz risk measures are always return risk measures. 
Their convexity properties are determined by the convexity properties of the function $\Phi$.\ 
AA-convexity has been studied in the seminal paper \cite{HG82}, while GG-convexity has been studied in \cite{ABL23}.
Summing up, we have the following:

\begin{itemize}
\item  $H_\Phi$ is AA-convex if and only if $\Phi$ is AA-convex
\item $H_\Phi$ is GA- and GG-convex if and only if $\Phi$ is GA-convex.
\end{itemize}

Indeed, Example~\ref{ex:Gmean} corresponds to $\Phi(x) = 1 + \log x$ and Example~\ref{ex:pnorm} corresponds to $\Phi(x) = x^p$, and both Orlicz functions are GA-convex and not AA-convex. 
\medskip

Similar to what we have done in Proposition~\ref{prop:GG-duality} for real functions of a real variable, we develop a duality theory for GG-convex risk measures.\ 
The first step is the definition of the GG-convex conjugate.\ 

\begin{definition} 
Let $\rho\colon \X_{\rm{log}}  \to [0,+\infty]$.\ 
The GG-convex conjugate of $\rho$ is denoted by $\rho^\di \colon \X^\ast_{\rm{log}} \to [0, +\infty]$ and given by
\[
\rho^\di(Y)=\sup_{X \in \X_{\rm{log}}} \left \{ \frac{\exp ( \E [\log Y \log X])}{\rho (X)} \right \}. 
\]
\end{definition}
The main properties of the GG-convex conjugation are similar to the scalar case analyzed in Section~\ref{sec:GG-dual-1} and will not be repeated. 
In order to have a suitable duality representation, it is necessary to have lower semicontinuity, which in this setting corresponds to a version of the Fatou property introduced in \cite{ABL23}. 
 
\begin{definition}
A risk measure $\rho \colon \X \to [0,+\infty]$ has the lower-bounded Fatou property if 
\[
X_n \overset{\p}{\to} X, \, 0 < a \leq \vert {X_n} \vert \leq b \, \, \p \text {-a.s.} \implies \rho(X)\leq \liminf_{n\to +\infty} \rho(X_n).
\]
\end{definition}

In the next proposition we give a general dual representation for GG-convex risk measures. 
\begin{proposition} 
\label{prop:GG-duality-2}
Let $\rho \colon \X_{\log} \to [0,\infty]$ be GG-convex and satisfy the lower-bounded Fatou property. Then 
\begin{equation}\label{dual-rep-2}
\rho (X)=\sup_{Y \in \X^\ast_{\rm{log}} } \left \{ \frac{\exp ( \E [\log Y \log X])}{\rho^\di (Y)} \right \}. 
\end{equation}
\noindent
Further, 
\begin{itemize}
\item [a)] $\rho$ is monotone if  $ Y \geq 1$ for each $Y \in \dom \rho^\di$
\item [b)] $\rho$ is positively homogeneous if $\E[\log Y] =1$ for each $Y \in \dom \rho^\di$
\item [c)] $\rho$ is normalized if $\inf_{Y \in \mathcal{X}^\ast_{\rm{log}} }  \rho^\di(Y) = 1$. 
\end{itemize}
\end{proposition}
It is interesting to compare equation \eqref{dual-rep-2} with the dual representation of GG-convex return risk measures given in Theorem~8 of \cite{ABL23}; see also \cite{LS13}.\ 
Letting $\Q_Y$ be the absolutely continuous signed measure on $(\Omega, \F)$ with Radon-Nikodym derivative
$
d\Q/d\p = \log Y,
$
and letting 
$
\alpha(\Q) = \left( \rho^\di (\exp (d\Q/d\p)) \right)^{-1},
$
equation \eqref{dual-rep-2} becomes
\begin{equation*}
\rho(X)=\sup_{\Q \ll \p} \{ \alpha(\Q)\exp\of{\E_{\Q}[\log X]} \},
\end{equation*}
which is the dual representation of a GG-convex risk measure as a worst-case weighted logarithmic certainty equivalent given in Theorem~8 of \cite{ABL23}, where $\alpha$ is a suitable multiplicative penalty function.  

\begin{example}
Let $\rho(X)=\Vert X \Vert _p$, with $p>0$.  
Then, with some calculations,
\[
\alpha(\Q)=\exp\of{-\frac{1}{p}H(\Q,\p)},
\]
where 
\[
H(\Q,\p)=
\begin{cases}
\E \left [ \frac{d\Q}{d\p} \log \frac {d\Q}{d\p} \right] \text { if } \Q \ll \p \\
+ \infty \text { otherwise}. 
\end{cases}
\]

Therefore, the $p$-norm has the following dual representation:
\begin{equation*}
    \Vert{X}\Vert_p=\sup_{\Q \ll \p} \exp \left (\E_\Q[\log(X)]-\frac{1}{p}H(\Q,\p) \right),
\end{equation*}
which can also be derived from the duality formula given in Example~4.34 of \cite{FS11},
\[
\frac{1}{\gamma}\log \E[\exp(\gamma Y)]=\sup_{\Q\in \mathbf{P}}\left(\E_\Q[Y]-\frac{1}{\gamma}H(\Q,\p)\right),
\]
letting $\gamma=p$ and $Y=\log(X)$.
\end{example}

\section{Stochastic orders and consistency results}\label{sec:GG-Order}
In this section, we introduce two stochastic orders based on GA-convex functions and on nondecreasing GA-convex functions.\ 
We start by recalling the definitions of the usual stochastic order $\leq_{\rm{st}}$, of the convex order $\leq_{\rm{cx}}$, and of the increasing convex order $\leq_{\rm{icx}}$.\ 
We refer e.g.,\ to \cite{MS02, SS07} for the properties of these usual stochastic orders and for a general exhaustive treatment of the theory of stochastic orders.
It follows from Lemma~\ref{lem:AM-GM} that the new increasing GA-convex order that we introduce is \textit{in between} $\leq_{\rm{st}}$ and $\leq_{\rm{icx}}$: it is weaker than $\leq_{\rm{st}}$ and stronger than $\leq_{\rm{icx}}$. 
\begin{definition}
Let $X,Y \colon (\Omega, \F, \p) \to \R$ be random variables.\ 
It holds that 
$X \leq_{\rm{st}} Y$, $X \leq_{\rm{cx}} Y$, $X \leq_{\rm{icx}} Y$ if
$
\E[f(X)] \leq \E[f(Y)],
$
for each $f \in \F_{\rm{st}}$, $f \in \F_{\rm{cx}}$, $f \in \F_{\rm{icx}}$, respectively, such that both expectations exist, where
\begin{align*}
\F_{\rm{st}}&: = \{f \colon \R \to \R \mid f \text { is nondecreasing} \}\\
\F_{\rm{cx}}&: = \{f \colon \R \to \R \mid f \text { is convex} \}\\
\F_{\rm{icx}}&: = \{f \colon \R \to \R \mid f \text { is nondecreasing and convex} \}.
\end{align*} 
\end{definition} 

We define similarly the GA-convex order and the increasing GA-convex order. 
\begin{definition} 
Let $X,Y \colon (\Omega, \F, \p) \to (0,+\infty) $ be positive random variables. 
It holds that
$X \leq_{\rm{GA-cx}} Y$, $X \leq_{\rm{GA-icx}} Y$ if 
$
\E[f(X)] \leq \E[f(Y)],
$
for each $f \in \F_{\rm{GA-cx}}$, $f \in \F_{\rm{GA-icx}}$, respectively, such that both expectations exist, where
\begin{align*}
\F_{\rm{GA-cx}}&: = \{f \colon \R \to \R \mid f \text { is GA-convex} \}\\
\F_{\rm{GA-icx}}&: = \{f \colon \R \to \R \mid f \text { is nondecreasing and GA-convex} \}.
\end{align*} 
\end{definition}
It is immediate to see that, since
\begin{align*}
f \in \F_{\rm{GA-cx}} &\iff f = g \circ \log, \text { with } g \in \F_{\rm{cx}} \\
f \in \F_{\rm{GA-icx}} &\iff f = g \circ \log, \text { with } g \in \F_{\rm{icx}}, 
\end{align*}
it follows that
\begin{align*}
X \leq_{\rm{GA-cx}}Y &\iff \log X \leq_{\rm{cx}} \log Y \\
X \leq_{\rm{GA-icx}}Y &\iff \log X \leq_{\rm{icx}} \log Y. 
\end{align*}

\begin{remark}
It is interesting to note that these orders can also be defined by means of GG-convex and nondecreasing GG-convex functions, since it holds that
$X \leq_{\rm{GA-cx}} Y$, $X \leq_{\rm{GA-icx}} Y$ if 
$
\G[f(X)] \leq \G[f(Y)],
$
for each $f \in \F_{\rm{GG-cx}}$, $f \in \F_{\rm{GG-icx}}$, respectively, such that both expectations exist, where
\begin{align*}
\F_{\rm{GG-cx}}&: = \{f \colon \R \to \R \mid f \text { is GG-convex} \}\\
\F_{\rm{GG-icx}}&: = \{f \colon \R \to \R \mid f \text { is nondecreasing and GG-convex} \}.
\end{align*} 
\end{remark}

For random variables on finite equiprobability spaces, the $\leq_{\rm{GA-cx}}$ order corresponds to the notion of multiplicative majorization or log-majorization introduced in \cite{W49} (see also \cite{N00, MOA11}). 
As is customary, we will write $F \leq_{\rm{GA-cx}} G$ or $F \leq_{\rm{GA-icx}} G$ if $X \sim F$ and $Y \sim G$ with $X \leq_{\rm{GA-cx}} Y$ or $X \leq_{\rm{GA-icx}} Y$.\ 

A few elementary necessary and sufficient conditions for these orders are given in the following lemma. 
Recall the notation for the number of sign changes of a generic function $f \colon \R \to \R$ as introduced e.g.,\ in \cite{SS07}, p.\ 10:
\[
S^-(f) = \sup S^-[f(x_1), \ldots, f(x_m)], 
\]
where $S^-[a_1, \ldots, a_m]$ is the number of sign changes in the sequence $a_1, \ldots, a_m$ and the supremum is taken over all $x_1 < \cdots < x_m$ and over all $m \in \mathbb{N}$. 

\begin{lemma}\label{lem:ord-1}
\begin{itemize}
\item []
\item [a)]  $X \leq_{\rm{GA-cx}} Y \Rightarrow \G(X) = \G(Y)$
\item [b)] $X \leq_{\rm{GA-icx}} Y \Rightarrow \G(X) \leq \G(Y)$

\item [c)] If $Y=XZ$ with $Z \perp X$ and $\G(Z)=1$ then $X \leq_{\rm{GA-cx}} Y$
\item [d)] Let $X \sim F$ and $Y\sim G$ with $\G(X)=\G(Y)$. If $S^-(G-F)=1$ with sign sequence $[+,-]$ then 
$X \leq_{\rm{GA-cx}} Y$. 
\end{itemize}
\end{lemma}

Many other properties of these orders can be derived from the corresponding properties of the $\leq_{\rm{cx}}$ and $\leq_{\rm{icx}}$ orders.\ 
The connection with the theory of GG-convex risk measures is given by the following general consistency results. 

\begin{proposition}\label{prop:consistency}
Let $\rho \colon \mathcal{X}_{\log} \to [0,\infty]$ be a law-invariant GG-convex risk measure.\ 
Then 
\[
X \leq_{\rm{GA-cx}} Y \Rightarrow \rho(X) \leq \rho (Y). 
\]
Further, if $\rho$ is also monotone, then 
\[
X \leq_{\rm{GA-icx}} Y \Rightarrow \rho(X) \leq \rho (Y). 
\]
\end{proposition}

\begin{remark}
For the special case of return risk measures defined on $\X_{\log}=L^\infty_{++}$, Proposition~\ref{prop:consistency} can also be obtained as a corollary of the following Kusuoka-like representation of GG-convex risk measures given in Theorem~11 of \cite{ABL23}:
\begin{equation*} 
	\rho(X) = \sup_{\mu\in\mathcal{M}_{1}([0,1])} \left \{ \beta(\mu)\exp\left(\int_{[0,1]}AV@R_{\lambda}(\log X)\mu(d \lambda)\right) \right \},
\end{equation*}
where $\mathcal{M}_1([0,1])$ is the set of probability measures with support in $[0,1]$, $\beta \colon \mathcal{M}_{1}([0,1]) \to [0,1]$ and 
\[
AV@R_\lambda (X) = 
\begin{cases}
\frac{1}{1-\lambda} \int_\lambda^1 q_\alpha(X) \, d \alpha, &\text { if } 0 \leq \lambda < 1, \\
\esssup(X), &\text { if } \lambda = 1,
\end{cases}
\]
with
\[
q_\alpha(X) = \inf \{ x \in \R \mid F(x) \geq \alpha \}.
\]
Indeed, it is well known that $AV@R_\lambda (X)$ is consistent with respect to the $\leq_{\rm{cx}}$ and the $\leq_{\rm{icx}}$ orders. 
\end{remark}

\section{Conclusion}\label{sec:con}
Quite surprisingly, considering the notion of GG-convexity opens up a rich, hitherto largely unexplored domain of research.\ 
In future work, extensions in several directions are conceivable.\ 
As an example, one can analyze GG-convexity in a multidimensional setting.\
This may lead to the definition of geometric conjugates on $d$-dimensional spaces, with $d>1$,
and to unveiling links to multivariate or systemic risk measures as well as to multivariate stochastic ordering.

\section{Appendix}

\subsection{Proof of Proposition~\ref{prop:GG-prop}}
We first prove c).\
Assume first that $\im f = (0, \infty)$.\ 
From the definition of GG-convexity it follows that
$
\log f(x^\lambda y^{1-\lambda}) \leq \lambda \log f(x) + (1-\lambda) \log f(y)
$
and setting $x=\exp(u)$, $y=\exp(v)$ gives 
\[
\log f(\exp(\lambda u + (1-\lambda)v)) \leq \lambda \log f(\exp(u)) + (1-\lambda) \log f(\exp(v)),
\]
that is, $g:=\log  \circ \, f \circ \exp$ is convex and finite-valued, so
$
f = \exp  \circ \, g  \circ  \log. 
$
More generally, the only case that leads to an indeterminate form is $f(x)=0$, $f(y)=\infty$. 
Setting by definition $\infty - \infty = \infty$ we recover the usual definition of convexity for extended-valued functions as in \cite{Z02}. 
To prove a), notice that from c) it holds that $f$ is GG-convex if and only if 
$
f = h  \circ  \log,
$
where $h$ is log-convex. 
Since the sum of log-convex functions is log-convex (see e.g.,\ \cite{RV73}, Theorem 13.F), the thesis follows. 
The case of products follows immediately from the definition. 
To prove b), as for the usual convexity it holds that
\[
\sup_{\alpha \in I} f_\alpha (x^\lambda y^{1-\lambda}) \leq \sup_{\alpha \in I} \{f^\lambda(x) \cdot f^{1-\lambda}(y)\} \leq [\sup_{\alpha \in I}f_\alpha(x)]^\lambda \cdot [\sup_{\alpha \in I}f_\alpha(y)]^{1-\lambda}.
\]
For d), since 
$
g(t)=\log (f(\exp(t)))
$
is convex from c), Jensen's inequality gives
\[
\log (f(\exp(\E[Y]))) \leq \E[\log (f(\exp(Y))],
\]
hence letting $Y=\log X$ and exponentiating both sides gives the thesis.
Finally, e) follows from c) and a direct computation, see e.g.,\ \cite{NP04} p.\ 87. 

\subsection{Proof of Proposition~\ref{prop:GG-conjugate}}
a) From
\[
f^{\di}(y) = \sup_{x>0} \left \{ \frac{x^{\log y}}{f(x)}  \right \} 
\]
it follows that $f^\di$ is GG-convex since it is the supremum of GG-affine functions (Proposition~\ref{prop:GG-prop}\, item c). 
It is also lower semicontinuous since $f^{\di}$ is the supremum of a family of continuous functions.\ 
b) follows immediately from the definition. 
To prove c), note that 
\begin{align*}
\sup_{\alpha \in I} (f_\alpha^\di) &= 
\sup_{\alpha \in I} \sup_{x>0}\left \{ \frac{\exp (\log x \log y)}{f_\alpha(x)} \right\} = 
\sup_{x>0}\sup_{\alpha \in I} \left \{ \frac{\exp (\log x \log y)}{f_\alpha(x)} \right\} \\
&= \sup_{x>0}\left \{ \frac{\exp (\log x \log y)}{\inf_{\alpha \in I} f_\alpha(x)} \right\} =
\left [\inf_{\alpha \in I} f_\alpha \right]^\di.
\end{align*}
To prove d), note that from b)
$
f_\alpha \leq \sup_\alpha f_\alpha \Rightarrow f_\alpha^\di \geq \left [\sup_\alpha f_\alpha \right]^\di
$
for each $\alpha$ and taking the infimum with respect to $\alpha$ gives the thesis.\ 
e) is immediate. 
To prove f), notice that
\begin{align*}
[f(Ax)]^\di(y) = \sup_{x>0} \left \{ \frac{y^{\log x}}{f(Ax)} \right \} = 
\sup_{z>0} \left \{ \frac{y^{(\log z - \log A)}}{f(z)} \right \} = f^\di(y) \cdot y ^{\log A}. 
\end{align*}
To prove g), 
\begin{align*}
[f(x^A)]^\di(y) = \sup_{x>0} \left \{ \frac{x^{\log y}}{f(x^A)} \right \} = 
\sup_{z>0} \left \{ \frac{z^{(\log y /A)}}{f(z)} \right \} = f^\di(y^{1/A}).
\end{align*}
Finally, to prove h), 
\begin{align*}
[f(x) \cdot x^A]^\di(y) = \sup_{x>0} \left \{ \frac{x^{\log y}}{f(x) \cdot x^A} \right \} = 
\sup_{x>0} \left \{ \frac{x^{(\log y- A)}}{f(x)} \right \} = f^\di(y/e^A).
\end{align*}

\subsection{Proof of Proposition~\ref{prop:GG-duality}}
From the definition of $f^\di$ it follows that
$
f(x) \cdot f^\di(y) \geq \exp (\log x \log y) 
$
for each $x,y >0$, so taking the supremum with respect to $y$ gives
\[
f(x)\geq \sup_{y>0} \frac{\exp (\log x \log y)}{f^\di(y)} = f^{\di \di}(x). 
\]

To prove the second part of the thesis, notice that if $f \colon (0,\infty) \to [0, \infty]$ satisfies
\[
f = \exp \circ \, g  \circ \log,
\]
then
\begin{align*}
f^\di(y) &= \sup_{x>0} \frac{\exp (\log x \log y)}{\exp \left (g(\log x) \right)}
= \sup_{z \in \R} \frac{\exp (z \log y)}{\exp \left( g(z) \right)} \\
&= \sup_{z \in \R}\exp \left ( z \log y - g(z) \right) = \exp \left \{ \sup_{z \in \R} \left (z \log y - g(z) \right) \right \} \\
&=\exp \circ \, g^\ast \circ \log(y),
\end{align*}
where 
$
g^\ast(y) = \sup_{x \in \R} \left  \{xy - g(x) \right\}
$
is the Fenchel conjugate of $g$.\ 
Since $f$ is GG-convex, it follows from Proposition~\ref{prop:GG-prop} that $g$ is convex. 
Since by assumption $f$ is proper and lower semicontinuous, it follows immediately that $g = \log \circ \, f  \circ \exp$
is also proper and lower semicontinuous, so $g^{\ast \ast} = g$ from the Fenchel-Moreau theorem. 
As a consequence, 
$
f^{\di \di} = \exp \circ \, g^{\ast \ast} \circ \log = \exp \circ \,  g \circ \log = f. 
$

\subsection{Proof of Theorem~\ref{th:GG-axiom} }
To prove the `if' part, assume that $T \colon \Gcal(0,\infty) \to \Gcal(0,\infty)$ has the form given in \eqref{Ti-ABC}.\
Recall from Proposition~\ref{prop:GG-duality} that $T(f)=f^\di$ is involutive and from Proposition~\ref{prop:GG-conjugate} that it is order-reversing.\ 
Let us consider the following cases. 
\begin{itemize}
\item [a)] Let $T(f)=Af^\di$, with $A>0$.\ 
Then $T$ is order-reversing and from Proposition~\ref{prop:GG-conjugate} item e) it is also involutive.
\item [b)] Let $(T(f))(x)=f^\di(x^C)$, for some $C \in \R$.\ 
Then $T$ is order-reversing and from Proposition~\ref{prop:GG-conjugate} item g) it is also involutive.
\item [c)] Let $(T(f))(x)=x^{\log B} f^\di(Bx)$.\ 
From Proposition~\ref{prop:GG-conjugate} items f) and h) it follows that
\[
[T(f)]^\di (x) = \left [ x^{\log B} f^\di(Bx)  \right]^\di = [f(Bx)]^\di (x/B) = f^\di(x/B) x^{-\log B},
\]
and
\[
T(T(f))(x) = x^{\log B } f^\di(B \cdot x/B) x^{-\log B} = f(x).
\]
\end{itemize}
Combining items a), b) and c) gives the `if' part. 
To prove the `only if' part, let
\[
\C (\R) = \left \{ f \colon \R \to \R \mid f \text{ is proper, convex and lsc} \right \}
\]
and for each $f \in \Gcal (0,\infty)$ we define $C(f) \in \C(\R)$ by
\[
C(f) = \log \circ \, f \circ \exp,
\]
where from Proposition~\ref{prop:GG-prop} it follows that $C \colon \Gcal (0,\infty) \to \C(\R)$ is one-to-one. 

For any $T \colon \Gcal(0,\infty) \to \Gcal (0,\infty)$, we define an associated transformation $\tilde{T} \colon \C(\R) \to \C(\R)$ by means of the following commutative diagram:
\[
\begin{tikzcd}
\mathcal{G}(0,\infty) \arrow[r,"T"] \arrow[d,"C"] &
\mathcal{G}(0,\infty) \arrow[d,"C"'] 
\\
\mathcal{C}(\R) \arrow[r, "\tilde{T}"] &
\mathcal{C}(\R)
\end{tikzcd}
\]
that is, $\Tilde{T} = C \circ T \circ C^{-1}$. 
$\Tilde{T}$ is one-to-one since both $C$ and $T$ are one-to-one. 
Further, $\Tilde{T}$ is involutive since
\[
(\Tilde{T} \circ \Tilde{T})(f) = (C \circ T \circ C^{-1} \circ C \circ T \circ C^{-1})(f)= (C \circ T \circ T \circ C^{-1})(f)=f,
\]
and is order-reversing since 
$
f \leq g \Rightarrow T(f) \geq T(g) \Rightarrow \Tilde{T}(f) \geq \Tilde{T}(g).
$
From Theorem~1 in \cite{AM09}, it follows that
\[
(\Tilde{T}(g))(x)=g^\ast(Cx+B)+Bx+A,
\]
with $A,B,C \in \R$, $A \neq 0$. 
Therefore, 
$
T = C^{-1} \circ \, \tilde{T} \circ C,
$
which gives \eqref{Ti-ABC}. 

\subsection{Proof of Proposition~\ref{prop:g-conv}}
Multiplicativity follows by a direct calculation, 
\begin{align*}
( f \gconv g)^\di(y) &= \sup_{x>0} \left \{ \frac{\exp(\log x \log y)}{(f \gconv g) (x) } \right \} =
\sup_{x>0} \left \{ \frac{\exp(\log x \log y)}{\inf_{x_1 \cdot x_2=x} f(x_1) g(x_2) } \right \} \\
&=\sup_{x>0} \left \{ \sup_{x_1 \cdot x_2=x}  \left \{ \frac{\exp(\log x \log y)}{ f(x_1) g(x_2) } \right \} \right \}  \\
&=\sup_{x_1>0, \, x_2>0} \left \{ \frac{\exp(\log x_1 \log y) \cdot \exp (\log x_2 \log y) }{ f(x_1) g(x_2) } \right \}  \\
& = \sup_{x_1 >0} \left \{ \frac{\exp(\log x_1 \log y)}{ f(x_1)} \right \} \cdot 
\sup_{x_2 >0} \left \{ \frac{\exp(\log x_2 \log y)}{ g(x_2)} \right \} = 
f^\di(y) \cdot g^\di (y). 
\end{align*}
Further, 
\begin{align*}
(C(f \otimes g ))(x) &= \log \left ( \inf_{x_1 \cdot x_2=e^x} (f(x_1)\cdot g(x_2)) \right) \\
&=\inf_{u_1 + u_2=x} \left ( C(f)(u_1) + C(g)(u_2)\right) =C(f) \oplus C(g), 
\end{align*}
and similarly 
\begin{align*}
(C^{-1}(f \oplus g ))(x) &= \exp \left ( \inf_{x_1+ x_2=\log x} (f(x_1)+ g(x_2)) \right) 
\\
&=\inf_{u_1 \cdot u_2=x} \left ( C^{-1}(f)(u_1) \cdot C^{-1}(g)(u_2)\right) =C^{-1}(f) \otimes C^{-1}(g).
\end{align*}

\subsection{Proof of Theorem~\ref{th:GG-axiom-2}}
To prove the `if' part assume that $T \colon \mathcal{G}(0,+\infty) \to \mathcal{G}(0,+\infty)$ has the form given in \eqref{Ti-A}. 
Then from Proposition~\ref{prop:g-conv} it follows immediately that
\[
T(f \gconv g) = (f \gconv g)^\di (x^A) = f^\di(x^A) \cdot g^\di (x^A) = T(f) \cdot T(g).
\]
Recall now from the proof of Theorem~\ref{th:GG-axiom} the construction of the transform $\Tilde{T}\colon \C(\R) \to \C(R)$ as 
$\Tilde{T} = C \circ T \circ C^{-1}$. 
As we have already shown, $\Tilde{T}$ is involutive. 
Further, $\Tilde{T}$ is additive with respect to inf-convolution since
\begin{align*}
\Tilde{T}(f \oplus g) &= C \circ T \circ C^{-1}(f \oplus g) = C \circ T \circ (C^{-1}(f) \otimes C^{-1}(g)) \\
&=C \circ (T(C^{-1}(f)) \cdot T(C^{-1}(g)))=\Tilde{T}(f) \oplus \Tilde{T}(g). 
\end{align*}

From Theorem~11 in \cite{AM07} it follows that
\[
(\Tilde{T}g)(x)=g^\ast(Bx),
\]
from which the thesis follows. 

\subsection{Proof of Lemma~\ref{lem:AM-GM-2} }
a), b) and c) follow from the AM-GM inequality. 
d) follows by noticing that if $\rho$ positively homogeneous, normalized and GA-convex, then
\[
\rho \left( \left [ \frac{X}{\rho(X)} \right]^\lambda \cdot \left[\frac{Y}{\rho(Y)}\right]^{1-\lambda}  \right) \leq \lambda \left [ \frac{X}{\rho(X)} \right] + (1-\lambda) \left[\frac{Y}{\rho(Y)}\right] = 1,
\]
which gives the thesis. 

\subsection{Proof of Proposition~\ref{prop:GG-duality-2}}
The proof is similar to the proof of Proposition~\ref{prop:GG-duality} and relies on the Fenchel duality theorem. 
The first step is to note that, if $\rho$ is GG-convex, then 
\[
\rho = \exp \circ \, \tilde{\rho} \circ \log,
\]
where $\tilde{\rho} \colon \X \to \R$ is convex (see \cite{BLR18}, Lemma~2). 
Then 
\begin{align*}
\rho^\di(Y) &= \sup_{X \in \X_{\log}} \frac{\exp \left ( \E [ \log X \log Y ] \right) }{\exp \left( \trho( \log X) \right)}=
\sup_{Z \in \X} \frac{ \exp \left ( \E[Z \log Y] \right) }{\exp (\trho(Z))} \\
&=\sup_{Z \in \X} \exp \left ( \{ \E[ Z \log Y]  - \trho (Z) \} \right ) \\
& = \exp \left ( \sup_{Z \in \X} \{ \E[ Z \log Y]  - \trho (Z) \} \right) = \exp \left ( \trho^\ast (\log Y) \right), 
\end{align*}
i.e.,\ we have shown that
\[
 \rho^\di = \exp \circ \, \trho^\ast \circ \log,
\]
where for $Y \in \X^\ast$
\[
\trho^\ast (Y) = \sup_{Z \in \X} \{ \E[ZY] - \rho (Z) \}
\]
is the Fenchel conjugate of $\trho$. 
Under the assumptions on $\rho$, it follows that $\trho$ is proper and has the Fatou property (see \cite{ABL23}, Lemma~6), so it is lower semicontinuous with respect to the $\sigma(\X, \X^\ast)$-topology and admits the representation
\[
\trho(X) = \sup_{Y \in \X^\ast} \{ \E[XY] - \trho^\ast (Y) \}. 
\]
As a consequence,
\begin{align*}
\rho (X) & = \exp \left ( \trho (\log X) \right) = \exp \left ( \sup_{Y \in \X^\ast} \{ \E[Y \log X] - \trho^\ast (Y) \} \right) \\
& = \sup_{Y \in \X^\ast} \exp \left (  \{ \E[Y \log X] - \log (\rho^\di( \exp ( Y ))) \}   \right ) \\
& = \sup_{Y \in \X^\ast} \exp \left (  \{ \E[ \log Y \log X] - \log (\rho^\di( Y )) \}   \right ) \\
& = \sup_{Y \in \X^\ast} \frac{\exp \left (   \E[ \log Y \log X]    \right )}{\rho^\di( Y )} . 
\end{align*}
To prove a), notice that $Y \geq 1 \Rightarrow \log Y \geq 0$, from which monotonicity follows. 
To prove b), notice that 
\[
\rho (\lambda X) = \sup_{Y \in \X^\ast_{\log}} \left \{ \frac{\exp(\E [\log Y (\log X + \log \lambda])}{\rho^\di(Y)}  \right \} = \lambda \rho(X).
\]

Finally, since
\[
\rho(1) = \sup_{Y \in \X^\ast_{\log}} \frac{1}{\rho^\di(Y)} =  \inf_{Y \in \X^\ast_{\log}} \rho^\di(Y) = 1,
\]
item c) follows. 

\subsection{Proof of Lemma~\ref{lem:ord-1}}
a) Since for each $a \in \R$ the function $x^a$ is GA-convex, we have
$$
X \leq_{\rm{GA-cx}} Y \Rightarrow \E[X^a] \leq \E[Y^a] \text { for each } a \in \R.
$$
As a consequence, for each $a>0$
it holds that
$\left (\E[X^a] \right)^{1/a} \leq  \left (\E[Y^a] \right)^{1/a}$
and for each $a<0$ it holds that
$
 \left (\E[X^a] \right)^{1/a} \geq  \left (\E[Y^a] \right)^{1/a}.
$
Since, as is well known, $$\G(X)=\lim_{a \to 0} \left ( \E[X^a] \right)^{1/a},$$ the thesis follows. 
The proof of b) is identical; here only the case $a \geq 0$ applies. \\
c) It holds that
\[
\log X \leq_{\rm{cx}} \log Y = \log X + \log Z,
\]
since $\log Z \perp \log X$ and $\E[\log Z]=\log \G(Z)=0$ (Theorem~3.A.34 in \cite{SS07}). \\
d)
It holds that
\[
\p(\log X \leq t) = F(e^t) \text { and } \p(\log Y \leq t) = G(e^t),
\]
so from the assumption that $F$ and $G$ cut in one point it follows that also the distribution functions of $\log X$ and $\log Y$ cut in one point, and since $\E[\log X] = \E[\log Y]$ the thesis follows. 

\subsection{Proof of Proposition~\ref{prop:consistency}}
If $\rho \colon \X_{\log} \to [0,+\infty]$ is GG-convex and law-invariant, then
\[
\rho = \exp \circ \, \trho \circ \log,
\]
where $\trho \colon \X \to \R$ is convex and law-invariant. 
From the results in \cite{BM06} and more generally in \cite{BKMMS21} it follows that $\trho$ is consistent with respect to the $\leq_{\rm{cx}}$ order. 
As a consequence, 
\[
\log X \leq_{\rm{cx}} \log Y \Rightarrow \rho(X) \leq \rho(Y).
\]
Similarly, if $\rho$ is also monotone, then $\trho$ is also monotone, so 
\[
\log X \leq_{\rm{icx}} \log Y \Rightarrow \rho(X) \leq \rho(Y).
\]


\begin{thebibliography}{99}

\bibitem {ADB19} \textsc{Agrawal, A., Diamond, S., Boyd, S.} (2019). 
Disciplined geometric programming.  
\textit{Optimization Letters} 13, 961-976.

%
%

\bibitem{AM07} \textsc{Artstein-Avidan, S., Milman, V.} (2007). 
A characterization of the concept of duality. 
\textit{Electronic Research Announcements in Mathematical Sciences} 14, 42-59. 

\bibitem{AM09} \textsc{Artstein-Avidan, S., Milman, V.} (2009). 
The concept of duality in convex analysis, and the characterization of the Legendre transform. 
\textit{Annals of Mathematics} 169, 661-674.

\bibitem{ABL23} \textsc{Aygün, M., Bellini, F., Laeven, R.J.A.} (2023). 
Elicitability of return risk measures. 
Working paper, \url{https://arxiv.org/abs/2302.13070v2}.

\bibitem{BM06} \textsc{B\"auerle, N., M\"uller, A.} (2006). 
Stochastic orders and risk measures: Consistency and bounds. \textit{Insurance: Mathematics and Economics} 38, 132-148. 

%
\bibitem{BRG08} \textsc{Bellini, F., Rosazza Gianin, E.} (2008).
On Haezendonck risk measures.
\textit{Journal of Banking and Finance} 32(6), 986–994.

\bibitem{BLR18} \textsc{Bellini, F., Laeven, R.J.A., Rosazza Gianin, E.} (2018). 
Robust return risk measures. 
\textit{Mathematics and Financial Economics} 12(1), 5-32.

\bibitem{BKMMS21} \textsc{Bellini, F., Koch-Medina, P., Munari, C., Svindland, G.} (2021). 
Law-invariant functionals on general spaces of random variables. \textit{SIAM Journal on Financial Mathematics} 12(1), 318-341.
%
%
%
%
\bibitem{FS11} \textsc{F\"ollmer, H., Schied, A.} (2011). Stochastic Finance. Third edition, De Gruyter, Berlin.
%
%
\bibitem{HG82} \textsc{Haezendonck, J., Goovaerts, M.J.} (1982).
A new premium calculation principle based on Orlicz norms.
\textit{Insurance: Mathematics and Economics} 1(1), 41-53.

\bibitem{HLP52} \textsc{Hardy, G., Littlewood, J.E., Polya, G.} (1952).
\textit{Inequalities}.
Cambridge Mathematical Library, 2nd Edition.

\bibitem{LS13} \textsc{Laeven, R.J.A., Stadje, M.A.} (2013). 
Entropy coherent and entropy convex measures of risk. 
\textit{Mathematics of Operations Research} 38, 265-293.

\bibitem{LRG22} \textsc{Laeven, R.J.A., Rosazza Gianin, E.} (2022). 
Quasi-logconvex measures of risk. 
Working paper, \url{https://arxiv.org/abs/2208.07694}.

\bibitem{MOA11} \textsc{Marshall, A.W., Olkin, I., Arnold, B.C.} (2011). 
Inequalities: Theory of Majorization and Its Applications. 
Springer Series in Statistics, New York. 

\bibitem{M28} \textsc{Montel, P.} (1928).
Sur les fonctions convexes et les fonctions sousharmoniques.
\textit{Journal de Math\'ematiques Pures et Appliqu\'ees} 9(7), 29-60.

\bibitem{MS02} \textsc{M\"uller, A., Stoyan, D.} (2002). 
Comparison Methods for Stochastic Models and Risks. 
Wiley Series in Probability and Statistics, New York. 

\bibitem{N00} \textsc{Niculescu, C.P.} (2000). 
Convexity according to the geometric mean. 
\textit{Math. Inequal. Appl.} 2, 155-167.

\bibitem{NP04} \textsc{Niculescu, C.P., Persson L.-E.} (2004). 
Convex Functions and Their Applications. 
Springer, New York. 

\bibitem{RV73} \textsc{Roberts, A.W., Varberg, D.E.} (1973). 
Convex Functions. 
Academic Press, New York and London.

\bibitem{R70} \textsc{Rockafellar, R.T.} (1970). 
Convex Analysis. 
Princeton University Press, Princeton.

\bibitem{SS07} \textsc{Shaked, M., Shantikumar, J.G.} (2007). 
Stochastic Orders. 
Springer Series in Statistics, New York.

\bibitem{W49} \textsc{Weyl, H.} (1949).
Inequalities between two kinds of eigenvalues of a linear transformation. 
\textit{Proc. Nat. Acad. Sci. USA} 35, 48-111. 

\bibitem{Z02} \textsc{Zalinescu, C.} (2002). 
Convex Analysis in General Vector Spaces. 
World Scientific, Singapore.
\end{thebibliography}
\end{document}